\documentclass[12pt, a4paper]{amsart}

\usepackage[utf8]{inputenc}
\usepackage[T1]{fontenc}
\usepackage{lmodern}
\usepackage{amsmath,amssymb} 
\usepackage{graphicx}
\usepackage{booktabs}
\usepackage{xurl}            
\usepackage[margin=1in]{geometry} 
\usepackage{amsthm} 
\usepackage{comment}
\usepackage{microtype} 
\usepackage[normalem]{ulem}

\usepackage{tcolorbox}
\usepackage{tikz}

\usepackage{hyperref}       

\usepackage[
  backend=biber,             
  citestyle=authoryear,      
  bibstyle=numeric,          
  maxcitenames=2,            
  maxbibnames=99             
]{biblatex}

\usepackage[dvipsnames]{xcolor}
\usepackage{enumitem}

\setlist[enumerate]{
  before=\vspace{0.5ex},
  after=\vspace{0.5ex},
  topsep=0.1ex,
  itemsep=0.5ex,
  parsep=0ex,
}

\setlist[itemize]{
  before=\vspace{0.5ex},
  after=\vspace{0.5ex},
  topsep=0.1ex,
  itemsep=1ex,
  parsep=0ex,
}

\makeatletter
\renewcommand{\subsection}{\@startsection{subsection}{2}{\z@}%
  {-3.25ex\@plus -1ex \@minus -.2ex}%
  {1.5ex \@plus .2ex}%
  {\normalfont\normalsize\bfseries}} 
\makeatother

\makeatletter
\renewcommand{\subsubsection}{\@startsection{subsubsection}{3}{\z@}%
  {-3.25ex\@plus -1ex \@minus -.2ex}%
  {1.5ex \@plus .2ex}%
  {\normalfont\normalsize\bfseries}} 
\makeatother

\makeatletter
\renewcommand{\paragraph}{\@startsection{paragraph}{4}{\z@}%
  {3.25ex \@plus 1ex \@minus .2ex}%
  {0.5ex}%   <--- KEY CHANGE: Use a POSITIVE value (or 0pt)
  {\normalfont\normalsize\bfseries}}
\makeatother

\definecolor{light-gray}{gray}{0.95}

\selectfont

\title[A Decentralized Frontier AI Architecture]{A Decentralized Frontier AI Architecture Based on Personal Instances, Synthetic Data, and Collective Context Synchronization}

\author[J. Małecki]{Jacek Małecki}
\address{Department of Mathematics, Faculty of Mathematics, Wrocław University of Science and Technology}
\email{jacek.malecki@pwr.edu.pl}

\author[A. Mathiesen-Ohman]{Alexander Mathiesen-Ohman}
\address{AMOTHO Research Institute, Vallsjön 20, 780 00 Rörbäcksnäs, Sweden}
\email{amo@amotho.com}

\author[K. Tworek]{Katarzyna Tworek}
\address{Department of Management Systems and Organizational Development, Faculty of Management, Wrocław University of Science and Technology}
\email{katarzyna.tworek@pwr.edu.pl}

\dedicatory{\today}

\keywords{distributed artificial intelligence, large language models, federated learning, collective context field, synthetic learning signals, distributed cognition, decentralized AI systems
}

\begin{document}

\begin{abstract}
Recent progress in artificial intelligence has been driven largely by the scaling of centralized large language models through increased parameters, datasets, and computational resources. While effective, this paradigm introduces structural constraints related to compute concentration, energy consumption, data availability, and governance. This paper proposes an alternative architectural approach through the \textbf{H3LIX Decentralized Frontier Model Architecture (DFMA)}, a distributed AI framework in which locally operating AI instances generate synthetic learning signals derived from reasoning processes and interactions. These signals are aggregated within a shared contextual substrate termed the \textbf{Collective Context Field (CCF)}, which conditions reasoning behavior across the network without requiring direct parameter synchronization. By enabling contextual signal propagation rather than centralized retraining at every iteration, the architecture can be designed to support privacy-preserving collective learning under explicit assumptions, while facilitating distributed sharing of learned abstractions. The system further integrates  \textbf{Energy-Adaptive Model Evolution}, aligning learning activities with renewable energy availability to support more sustainable AI infrastructure. Conceptually, the architecture reframes artificial intelligence as a \textbf{distributed cognitive system} analogous to biological neural networks, in which intelligence emerges from the interaction of many locally adaptive agents within a shared contextual environment. Together, these mechanisms suggest a new scaling pathway for artificial intelligence systems based on distributed contextual learning and collective experience accumulation.
\end{abstract}
%\footnotetext[1]{\textbf{Alexander Mathiesen-Ohman, Provisional Utility Patent Pending, no. 63/910,500}}

\maketitle

\section{Introduction}
\subsection*{The Limits of the Centralized Large Language Model Paradigm}

The rapid progress of artificial intelligence in recent years has been driven largely by the development of \textbf{large language models (LLMs)}trained on massive datasets using increasingly large computational resources. Empirical scaling studies have demonstrated that model capability improves predictably with increases in model parameters, training data, and computational power (\cite{KaplanEtAl2020}). Subsequent work has refined these scaling relationships, showing that model performance is closely tied to compute-optimal training regimes and increasingly large training clusters (\cite{HoffmannEtAl2022} ). As a result, contemporary frontier AI systems rely on centralized architectures in which large neural networks are trained using hyperscale compute infrastructure and deployed through cloud-based inference services.

\medskip

\noindent This centralized paradigm has enabled major advances in natural language processing, reasoning, and generative capabilities. However, it also introduces several structural constraints that are becoming increasingly significant as model sizes continue to grow. First, the computational resources required to train frontier models have increased dramatically, concentrating AI development within a small number of organizations capable of operating large-scale training clusters. Training frontier models can require very large accelerator clusters and extended runtimes, creating substantial financial and infrastructure barriers (\cite{HoffmannEtAl2022}). If precise scales are needed, cite a source that explicitly reports cluster sizes and training durations; otherwise remove specific numbers. Second, the energy demands of such training runs are substantial, raising both economic and environmental concerns as artificial intelligence systems scale further (\cite{StrubellGaneshMcCallum2019}). Third, the availability of high-quality training data is increasingly constrained. Training frontier models can require very large accelerator clusters and extended runtimes, creating substantial financial and infrastructure barriers (\cite{HoffmannEtAl2022}). If precise scales are needed, cite a source that explicitly reports cluster sizes and training durations; otherwise remove specific numbers.

\medskip

\noindent In addition to these computational and data constraints, centralized LLM architectures present broader challenges related to governance, personalization, and user autonomy. Because model training and inference are typically centralized within cloud infrastructures, user interactions are often aggregated into shared datasets, raising concerns regarding privacy and data ownership. Furthermore, centralized models struggle to maintain persistent user identity and contextual continuity across interactions. While retrieval- and tool-augmented systems externalize important functions (\cite{LewisEtAl2020} ; \cite{SchickEtAl2023} ; \cite{YaoEtAl2023}), many deployments still treat the base language model as the primary reasoning component, even as memory and control are increasingly managed by surrounding runtimes (e.g., MemGPT; \cite{PackerEtAl2023MemGPT}).

\medskip

\noindent One prominent research direction addressing these concerns is \textbf{federated learning}, which allows distributed devices to train models collaboratively without sharing raw user data (\cite{McMahanEtAl2017}; \cite{KairouzEtAl2021}). In federated learning systems, model updates computed locally on client devices are aggregated to produce a shared global model. This approach improves privacy by keeping sensitive data on local devices while still enabling collective learning across distributed participants. Many federated learning systems are parameter-centric (e.g., FedAvg-style aggregation; \cite{McMahanEtAl2017}; \cite{KairouzEtAl2021}), although there are also non-parameter variants that transfer knowledge via outputs or shared prompts (e.g., federated distillation such as FedMD; \cite{LiWang2019FedMD}). H3LIX emphasizes synchronizing structured contextual artifacts (operators/priors/policies/curricula) rather than requiring global parameter convergence.

\medskip

\noindent Recent advances in parameter-efficient training techniques, such as low-rank adaptation (LoRA), have enabled more flexible distributed model adaptation by allowing smaller parameter modules to be updated independently (\cite{HuEtAl2021} ). These techniques have been applied to federated settings to support heterogeneous models and reduce communication overhead (\cite{YiEtAl2023}). Nevertheless, such approaches remain fundamentally grounded in the assumption that collective intelligence emerges through \textbf{parameter synchronization across distributed nodes}.

\medskip

\noindent This paper explores an alternative architectural paradigm. Instead of treating model parameters as the primary object of synchronization, we propose a decentralized AI architecture in which \textbf{collective learning emerges through the propagation of contextual reasoning signals across a distributed cognitive network}. The architecture, referred to as the \textbf{H3LIX Decentralized Frontier Model Architecture (DFMA)}, is composed of many personal AI instances operating locally on distributed devices. These instances generate synthetic learning signals derived from reasoning outcomes, simulations, and interaction patterns. Rather than aggregating model parameters, these signals are distilled into structured learning artifacts and synchronized across nodes through a distributed protocol.

\medskip

\noindent A central component of the architecture is the \textbf{Collective Context Field (CCF)}, a distributed layer of shared reasoning priors derived from aggregated learning signals across the network. Instead of updating a single global model through gradient descent, the Collective Context Field conditions the reasoning behavior of individual nodes by providing contextual operators, policy adjustments, and reasoning priors. In this way, improvements discovered by any participating node can propagate across the network without requiring centralized retraining of a monolithic model.

\medskip

\noindent Beyond distributed learning, the architecture introduces two additional structural innovations. First, the reasoning engine (e.g., a large language model) is separated from higher-order cognitive governance layers responsible for identity continuity, knowledge validation, and execution authority. This architectural separation allows reasoning models to be updated or replaced without disrupting long-term contextual continuity. Second, the system incorporates \textbf{energy-adaptive model evolution}, in which distributed learning activity is dynamically scheduled according to renewable energy availability and grid conditions. This mechanism allows AI training and synchronization processes to align with energy infrastructure constraints, potentially reducing the environmental footprint of large-scale AI development.

\medskip

\noindent Taken together, these mechanisms introduce a new architectural paradigm for artificial intelligence systems in which intelligence emerges not from a single centralized model but from the interaction of many locally adaptive agents within a shared contextual environment. Conceptually, this architecture resembles distributed cognitive processes observed in biological neural systems, where learning occurs through local adaptation combined with global modulatory signals that influence network-wide behavior (\cite{AstonJonesCohen2005}). Similarly, the generation of synthetic reasoning signals within local AI instances parallels mechanisms of hippocampal replay, in which biological neural systems simulate past experiences to strengthen learning and explore alternative behavioral trajectories (\cite{WilsonMcNaughton1994}; \cite{FosterWilson2006}).

\medskip

\noindent The present work develops this architectural paradigm and formalizes the mechanisms through which decentralized AI instances can generate, synchronize, and integrate synthetic learning signals through a Collective Context Field. By shifting the focus of distributed learning from parameter synchronization to contextual signal propagation, the architecture provides a potential pathway toward scalable AI systems capable of continuous collective improvement while preserving privacy, autonomy, and energy efficiency.

\section{Background and Related Literature}
\subsection{Centralized Large Language Model Architectures}
The rapid advancement of modern artificial intelligence has been largely driven by the development of large-scale neural language models trained using centralized computing infrastructures. Early empirical studies demonstrated that model performance scales predictably with increases in model parameters, training data, and computational resources (\cite{KaplanEtAl2020}). Subsequent work has refined these relationships, demonstrating that optimal performance emerges when models are trained using compute-efficient scaling regimes that balance dataset size, parameter count, and compute budget (\cite{HoffmannEtAl2022}).

\medskip

\noindent These findings have encouraged the development of increasingly large neural models trained on massive centralized datasets using hyperscale compute clusters. Frontier models such as GPT-family systems, PaLM, and other large transformer architectures are typically trained on centralized infrastructure containing thousands of specialized accelerators operating for extended periods. While this paradigm has produced significant improvements in language understanding, reasoning, and generative capabilities, it has also introduced structural constraints that are increasingly recognized in the literature.

\medskip

\noindent First, centralized training requires substantial computational resources, concentrating AI development within a small number of organizations capable of operating hyperscale infrastructure. Second, the energy demands associated with large-scale training runs have become significant. Studies examining the environmental impact of deep learning have shown that training large neural models may consume substantial amounts of energy and produce measurable carbon emissions, raising concerns regarding the long-term sustainability of current scaling strategies (\cite{StrubellGaneshMcCallum2019}).

\medskip

\noindent Third, the continued scaling of language models is increasingly constrained by the availability of high-quality training data. As large internet-scale corpora are progressively exhausted, recent research has emphasized the growing importance of synthetic data generation and alternative learning architectures capable of extracting more knowledge from existing data (\cite{KaplanEtAl2020}; \cite{HoffmannEtAl2022}). Together, these factors have motivated increasing interest in alternative AI architectures that can maintain or improve model capability while reducing dependence on centralized infrastructure.

\subsection{Federated Learning and Distributed Machine Learning}
One of the most prominent approaches for addressing privacy and decentralization concerns in machine learning is \textbf{federated learning (FL)}. Federated learning enables multiple distributed devices or organizations to collaboratively train a shared model while keeping the underlying training data local (\cite{McMahanEtAl2017}). Instead of transferring raw data to a central server, participating clients compute model updates locally and transmit parameter updates or gradients to a coordinating server, which aggregates these updates to produce a new global model.

\medskip

\noindent Federated learning has received substantial attention as a mechanism for enabling collaborative machine learning while preserving privacy and minimizing data transfer. A comprehensive survey of federated learning research highlights the wide range of challenges associated with this paradigm, including communication efficiency, heterogeneous client data distributions, privacy guarantees, and system robustness (\cite{KairouzEtAl2021}). In federated learning systems, multiple clients collaboratively train a shared model under the orchestration of a central server while maintaining decentralized datasets, thereby reducing many of the privacy risks associated with centralized machine learning pipelines.

\medskip

\noindent Despite these advantages, many widely deployed federated learning architectures remain parameter-centric (e.g., gradient/weight aggregation; \cite{McMahanEtAl2017}; \cite{KairouzEtAl2021}), although related lines of work reduce parameter exchange via knowledge transfer (e.g., federated distillation such as FedMD; \cite{LiWang2019FedMD}) or by synchronizing prompt/context objects (e.g., federated prompt learning). The core objective of most FL systems is to optimize a shared global model whose parameters are iteratively updated through distributed gradient aggregation. Even recent advances that incorporate parameter-efficient fine-tuning techniques, such as low-rank adaptation (LoRA), remain focused on synchronizing subsets of model parameters across participating nodes (\cite{HuEtAl2021}). Similarly, personalized federated learning approaches attempt to accommodate heterogeneous client environments by allowing local model customization while still relying on global parameter aggregation (\cite{YiEtAl2023}).

\medskip

\noindent These approaches have achieved considerable success in enabling privacy-preserving distributed learning, but they retain the central assumption that \textbf{model parameters represent the primary locus of intelligence}. As a result, improvements discovered by individual nodes must ultimately be translated into parameter updates that are synchronized across the network. This design creates scalability challenges related to communication cost, convergence stability, and model coordination in highly heterogeneous environments (\cite{KairouzEtAl2021}).

\medskip

\noindent Consequently, recent research has begun exploring alternative distributed learning paradigms that move beyond strict parameter synchronization, including peer-to-peer learning architectures, decentralized optimization algorithms, and hybrid distributed learning frameworks. However, most of these approaches still assume that collective learning ultimately converges toward a single shared model.

\subsection{Retrieval-Augmented and Agent-Based LLM Systems}
In parallel with the development of federated learning, researchers have explored various architectural extensions designed to augment the capabilities of large language models. One widely studied approach is \textbf{retrieval-augmented generation (RAG)}, which combines neural language models with external knowledge retrieval systems (\cite{LewisEtAl2020}). In RAG architectures, a neural retriever identifies relevant documents from an external knowledge corpus, and the retrieved information is incorporated into the model’s reasoning process. This approach allows LLMs to access large knowledge bases without requiring that all information be encoded directly within model parameters.

\medskip

\noindent More recent work has explored the integration of language models with external tools and action frameworks. For example, the ReAct framework combines reasoning and acting in language models by enabling models to generate intermediate reasoning steps that guide tool usage (\cite{YaoEtAl2023}). Similarly, the Toolformer framework demonstrates that language models can learn to invoke external APIs and computational tools autonomously (\cite{SchickEtAl2023}). These approaches significantly expand the capabilities of LLM-based systems by enabling models to interact with external knowledge sources and computational resources.

\medskip

\noindent Despite these advances, most agent-based or retrieval-augmented architectures remain built around the assumption that the language model itself serves as the primary reasoning engine within the system. External modules such as retrieval systems, memory stores, or tool interfaces typically function as extensions to the underlying model rather than as independent cognitive layers. In many systems the base model remains central to reasoning, but system capability can also improve substantially through better retrieval, memory management, tool use, and control policies (\cite{LewisEtAl2020}; \cite{YaoEtAl2023}; see also MemGPT, \cite{PackerEtAl2023MemGPT}). H3LIX emphasizes making these surrounding layers first-class and persistently synchronized at the system level.

\medskip

\noindent This limitation has motivated increasing interest in \textbf{system-level AI architectures} that treat language models as components within broader cognitive systems rather than as the sole source of intelligence. Such architectures aim to integrate reasoning engines with additional layers responsible for memory management, knowledge verification, decision governance, and long-term contextual continuity.

\subsection{Distributed Cognition and Biological Learning Systems}

\noindent Beyond machine learning research, insights from neuroscience and cognitive science provide valuable perspectives on distributed learning systems. Biological neural systems operate through highly distributed learning mechanisms in which local adaptation is combined with global modulatory signals that influence network-wide behavior. For example, neuromodulatory systems such as the locus coeruleus–norepinephrine system regulate cognitive processes across large regions of the brain, adjusting neural gain and influencing attention, learning, and decision-making (\cite{AstonJonesCohen2005}).

\medskip

\noindent Another important biological mechanism is \textbf{hippocampal replay}, in which neural activity patterns associated with past experiences are reactivated during sleep or rest. These replay events allow biological neural systems to consolidate learning and explore alternative behavioral trajectories, effectively generating synthetic experiences that strengthen memory and learning processes (\cite{WilsonMcNaughton1994}; \cite{FosterWilson2006}). Such mechanisms demonstrate that biological intelligence does not rely solely on direct experience but also on internally generated simulations that enable efficient learning.

\medskip

\noindent These principles have influenced research in reinforcement learning, generative modeling, and self-play systems, where synthetic experiences are used to accelerate learning. However, most current artificial intelligence architectures still rely heavily on centralized training procedures and static neural models rather than continuous distributed learning across networks of agents.

\medskip

\noindent The emerging literature on \textbf{distributed cognition} suggests that intelligence can arise from the coordinated activity of multiple interacting agents operating within a shared informational environment. Rather than being localized within a single computational entity, cognition may emerge from the dynamic interaction of multiple subsystems that exchange information and coordinate behavior through shared signals.

\medskip

\noindent These insights provide conceptual motivation for exploring AI architectures in which intelligence emerges from networks of locally adaptive agents operating within shared contextual environments. Such architectures may enable forms of collective intelligence that differ fundamentally from those produced by centralized neural models.

\section{Conceptual Foundations of the H3LIX Architecture}

\noindent The H3LIX architecture proposes a decentralized framework for artificial intelligence in which intelligence emerges from the interaction of multiple locally operating AI instances rather than from a single centralized model. This section introduces the conceptual foundations underlying the architecture and describes the core design principles that enable distributed learning, contextual synchronization, and persistent human–AI interaction.

\medskip

\noindent The architecture is based on five foundational concepts: \textbf{personal AI instances, separation of reasoning and governance, synthetic learning signals, decentralized knowledge synchronization, and the Collective Context Field}. Together these elements form the conceptual basis of the H3LIX Decentralized Frontier Model Architecture (DFMA), which aims to support scalable distributed intelligence while preserving user autonomy and data privacy.

\subsection{Personal AI Instances}

\noindent At the core of the H3LIX architecture is the concept of the \textbf{personal AI instance}, a locally operating AI system associated with an individual user or organizational node. Each personal instance maintains persistent contextual information about its environment and interactions, enabling long-term continuity of reasoning and personalization.

\medskip

\noindent In contrast to conventional cloud-based AI systems in which user interactions are processed by a centralized model, personal AI instances operate locally on user devices or edge computing nodes. These instances host reasoning models capable of conversational interaction, contextual synthesis, and problem-solving while maintaining local records of user interactions and preferences. By storing contextual information locally, the system enables persistent identity continuity without requiring centralized storage of user data.
The concept of personal AI instances aligns with broader trends toward \textbf{edge computing and distributed AI}, where computational intelligence is increasingly deployed closer to the source of data generation. Advances in edge hardware and model compression techniques have made it feasible for increasingly capable neural models to operate locally on consumer devices and distributed computing nodes (\cite{Satyanarayanan2017}). Such architectures can reduce latency, enhance privacy, and enable more personalized AI interactions.

\medskip

\noindent Within the H3LIX architecture, personal instances function as autonomous cognitive agents capable of generating learning signals and participating in collective knowledge formation. Each node contributes to the overall intelligence of the network while retaining control over local data and contextual information.

\subsection{Separation of Reasoning and Cognitive Governance}

\noindent A second foundational principle of the H3LIX architecture is the \textbf{separation between reasoning processes and higher-order governance mechanisms}. Contemporary large language model systems often combine reasoning, knowledge representation, contextual memory, and decision-making authority within a single neural architecture. While this integration simplifies system design, it also introduces challenges related to transparency, reliability, and governance.

\medskip

\noindent The H3LIX architecture separates these functions into distinct conceptual layers. The \textbf{reasoning engine}, typically implemented as a large language model or similar neural architecture, is responsible for generating responses, synthesizing information, and performing contextual reasoning. Surrounding the reasoning engine are several governance layers responsible for maintaining identity continuity, verifying knowledge claims, and regulating execution authority.

\medskip

\noindent The \textbf{identity layer} maintains persistent representations of users or system entities, enabling long-term contextual continuity across interactions. The \textbf{evidence layer}tracks the provenance and validation status of information used in reasoning processes, helping to reduce hallucination and misinformation risks. Finally, the \textbf{execution layer} governs interactions with external systems and real-world actions, ensuring that irreversible operations occur under appropriate constraints.

\medskip

\noindent Separating reasoning from governance mechanisms provides several advantages. First, it allows reasoning engines to be updated or replaced without disrupting the continuity of system identity or memory. Second, it improves transparency by isolating the decision-making process from the underlying reasoning model. Third, it provides a foundation for integrating verification mechanisms that help ensure reliability and safety in AI decision-making.

\medskip

\noindent This architectural separation reflects emerging perspectives in AI systems design that treat language models as components within broader cognitive infrastructures rather than as standalone intelligent entities (\cite{YaoEtAl2023}).

\subsection{Synthetic Learning Signals}

\noindent Another core concept underlying the H3LIX architecture is the use of \textbf{synthetic learning signals} generated by individual AI instances during interaction and reasoning processes. Traditional machine learning systems rely primarily on large static datasets collected from external sources. However, the availability of high-quality training data is increasingly constrained, motivating the exploration of synthetic data generation and self-supervised learning strategies.

\medskip

\noindent Within the H3LIX architecture, personal AI instances generate learning signals derived from reasoning outcomes, interaction feedback, and simulated reasoning trajectories. These signals may include corrected reasoning traces, distilled instruction-response pairs, counterfactual reasoning paths, and other forms of structured knowledge abstraction. Rather than representing raw interaction data, these signals encode \textbf{learning-relevant patterns extracted from reasoning processes}.

\medskip

\noindent Synthetic learning signals play a critical role in enabling continuous improvement of the distributed AI network without requiring centralized data collection. Because these signals are derived from abstractions rather than raw logs, they can be shared with reduced privacy risk when combined with explicit safeguards such as filtering, secure aggregation, and/or differential privacy (\cite{BonawitzEtAl2017}; \cite{KairouzEtAl2021}).

\medskip

\noindent The use of synthetic experiences for learning has precedent in both biological and artificial systems. In biological cognition, mechanisms such as hippocampal replay allow neural systems to simulate past experiences and hypothetical scenarios to strengthen learning (\cite{WilsonMcNaughton1994}; \cite{FosterWilson2006}). In artificial intelligence, self-play systems and simulated experience generation have been used to accelerate learning in reinforcement learning environments.
The H3LIX architecture extends these ideas to distributed AI systems by enabling each node to generate synthetic learning artifacts that contribute to collective intelligence formation.

\subsection{Decentralized Knowledge Synchronization}

\noindent To enable collective learning across distributed AI instances, the architecture introduces a mechanism for \textbf{decentralized knowledge synchronization}. Instead of sharing raw data or complete model parameters, personal instances periodically transmit distilled learning artifacts derived from their synthetic learning signals.

\medskip

\noindent These artifacts may include structured reasoning traces, knowledge abstractions, policy adjustments, or other forms of contextual learning signals. Because these artifacts represent distilled knowledge rather than raw data, they can be shared across nodes while preserving privacy.

\medskip

\noindent Distributed synchronization protocols aggregate these artifacts using privacy-preserving techniques such as secure aggregation and differential privacy mechanisms (\cite{BonawitzEtAl2017}). Aggregated signals are then incorporated into a shared contextual environment that influences the reasoning behavior of participating nodes.

\medskip

\noindent This approach differs from conventional federated learning architectures, which aggregate gradient updates to update a shared model. In the H3LIX architecture, the shared object of synchronization is not a model parameter vector but a \textbf{collection of contextual reasoning signals}. This distinction shifts the focus of distributed learning from parameter convergence to knowledge propagation across the network.

\subsection{Collective Context Field}

The central conceptual innovation of the H3LIX architecture is the \textbf{Collective Context Field (CCF)}, a distributed contextual layer that aggregates learning signals from across the network and conditions the reasoning processes of individual nodes.

\medskip

\noindent The Collective Context Field can be understood as a shared informational substrate that encodes reasoning priors derived from the aggregated experiences of participating nodes. Each personal AI instance contributes synthetic learning signals to the field and receives contextual updates that influence its subsequent reasoning behavior.

\medskip

\noindent Unlike conventional machine learning systems that update model parameters through centralized training, the CCF enables knowledge propagation through contextual conditioning. Improvements discovered by individual nodes can influence the reasoning behavior of other nodes through updates to the shared context field, even if the underlying model parameters remain unchanged.

\medskip

\noindent Conceptually, the Collective Context Field resembles global modulatory systems in biological cognition that influence neural processing across distributed networks. Neuromodulatory mechanisms such as dopamine and norepinephrine signaling regulate learning and attention across large regions of the brain without directly encoding specific information (\cite{AstonJonesCohen2005}). Similarly, the CCF influences reasoning processes by adjusting contextual priors rather than directly modifying model weights.

\begin{figure}[h!]
    \centering
    \includegraphics[width=\linewidth]{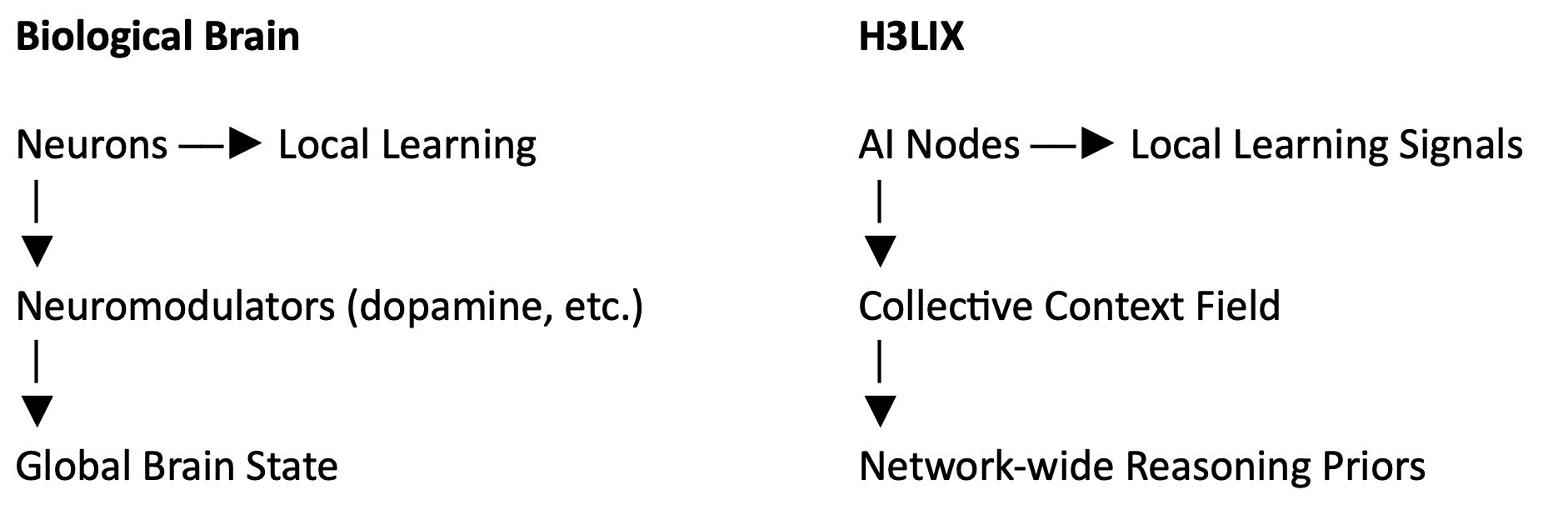}
    \caption{Comparison of H3LIX Architecture and Biological Brain}
\end{figure}

\medskip

\noindent This design allows the H3LIX network to support \textbf{continuous collective learning without requiring centralized retraining cycles}, enabling distributed intelligence to evolve through the interaction of many locally adaptive agents operating within a shared contextual environment.

\section{System Architecture of the H3LIX Network}

\noindent Building upon the conceptual foundations introduced in the previous section, the H3LIX architecture defines a distributed computational framework in which artificial intelligence emerges from the coordinated activity of multiple locally operating AI instances. This framework, referred to as the \textbf{Decentralized Frontier Model Architecture (DFMA)}, organizes the system into a layered structure that governs how reasoning, learning, synchronization, and contextual conditioning occur across the network.

\medskip

\noindent Unlike conventional large language model deployments that rely on centralized model hosting and inference services, the H3LIX architecture distributes intelligence across a network of personal AI instances that operate locally while exchanging structured learning signals through a decentralized synchronization protocol. The system architecture consists of four primary layers: the \textbf{Personal Instance Layer}, the \textbf{Synthetic Learning Layer}, the \textbf{Synchronization Layer}, and the \textbf{Collective Context Layer}. Together these layers enable continuous distributed learning while maintaining privacy, scalability, and resilience (Figure 1).

\begin{figure}[h!]
    \centering
    \includegraphics[width=\linewidth]{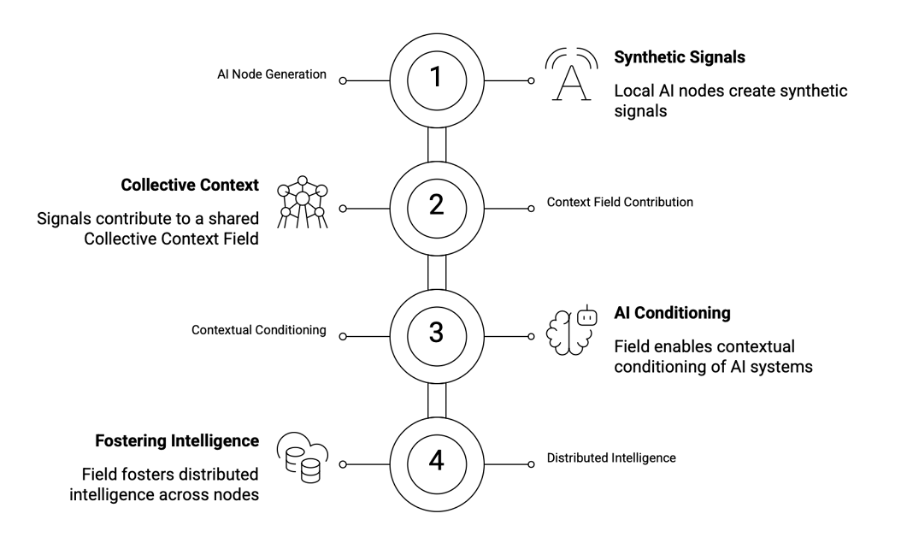}
    \caption{H3LIX Architectural Layer}
\end{figure}

\subsection{Personal Instance Layer}

The \textbf{Personal Instance Layer} forms the foundational computational layer of the H3LIX network. Each participating node operates a personal AI instance capable of performing reasoning, conversational interaction, and contextual synthesis. These instances typically host a compact reasoning model or adapter-based neural architecture capable of processing natural language inputs and generating responses within the context of local interactions.

\medskip

\noindent Each personal instance maintains several forms of local state, including interaction histories, contextual memory, and identity representations associated with the user or organizational entity operating the node. By maintaining this contextual information locally, the architecture allows each node to develop long-term contextual awareness while preserving user privacy. Unlike centralized AI systems in which user interactions are aggregated within shared cloud-based infrastructures, personal instances store and process sensitive data locally.

\medskip

\noindent In addition to providing conversational and reasoning capabilities, personal instances also function as \textbf{learning nodes} within the distributed network. During interactions with users or external systems, these nodes generate internal learning signals derived from reasoning outcomes and interaction feedback. These signals serve as the basis for distributed learning across the network.

\medskip

\noindent The personal instance layer therefore fulfills two primary roles. First, it acts as the primary interface between users and the AI system, enabling persistent and personalized interactions. Second, it serves as the local computational environment in which reasoning events generate learning artifacts that contribute to the evolution of the distributed network.

\subsection{Synthetic Learning Layer}

\noindent The \textbf{Synthetic Learning Layer} governs the generation of structured learning signals derived from reasoning processes occurring within personal AI instances. Rather than relying exclusively on external datasets or centralized training pipelines, the H3LIX architecture enables nodes to generate learning artifacts through self-reflection, simulation, and collaborative reasoning exchanges.

\medskip

\noindent During reasoning processes, personal AI instances evaluate the outcomes of their reasoning steps and extract learning-relevant patterns. These patterns may include corrected reasoning traces, validated instruction-response pairs, policy adjustments, and counterfactual reasoning trajectories. Instead of recording raw interaction logs, the system converts these patterns into \textbf{synthetic learning artifacts} that represent generalized knowledge abstractions derived from the reasoning process.

\medskip

\noindent Several mechanisms can generate such artifacts. For example, nodes may simulate alternative reasoning paths to evaluate potential improvements in problem-solving strategies. They may also engage in collaborative reasoning exchanges with other nodes to explore multiple perspectives on a given problem. Through these processes, nodes generate synthetic learning signals that capture improvements in reasoning behavior without revealing sensitive user information.

\medskip

\noindent This synthetic learning approach aligns with broader trends in artificial intelligence research emphasizing the role of self-generated experiences in accelerating learning processes. In reinforcement learning systems, self-play mechanisms have been shown to produce substantial improvements in agent performance without requiring external datasets. Similarly, biological neural systems employ mechanisms such as hippocampal replay to simulate past experiences and strengthen learning (\cite{WilsonMcNaughton1994}; \cite{FosterWilson2006}).
Within the H3LIX architecture, synthetic learning artifacts produced by the Synthetic Learning Layer become the primary units of distributed learning exchanged between nodes.

\subsection{Synchronization Layer}

\noindent The \textbf{Synchronization Layer} coordinates the exchange and aggregation of learning artifacts across the distributed network. Rather than transmitting raw data or full model updates, participating nodes periodically transmit distilled learning artifacts generated by the Synthetic Learning Layer.

\medskip

\noindent These artifacts are exchanged through a distributed synchronization protocol designed to preserve privacy and maintain system scalability. The synchronization process consists of several stages. First, nodes generate synthetic learning artifacts based on local reasoning events. Second, these artifacts undergo validation and filtering procedures designed to remove inconsistent or low-confidence signals. Third, validated artifacts are transmitted to aggregation nodes responsible for combining signals from multiple participants.

\medskip

\noindent To ensure privacy preservation, synchronization protocols may incorporate techniques such as secure aggregation and differential privacy mechanisms (\cite{BonawitzEtAl2017}). These mechanisms ensure that individual contributions cannot be reverse-engineered or traced back to specific users or interactions. As a result, nodes can contribute learning signals to the network without exposing raw interaction data.

\medskip

\noindent Importantly, the synchronization layer does not attempt to merge parameter updates into a shared global model. Instead, it aggregates learning artifacts into a distributed contextual representation that influences reasoning behavior across the network. This design distinguishes the H3LIX architecture from conventional federated learning systems, where synchronization typically aims to converge toward a shared parameterized model.

\subsection{Collective Context Layer}

\noindent The \textbf{Collective Context Layer} forms the highest-level layer of the H3LIX architecture and represents the central mechanism through which distributed intelligence emerges. Aggregated learning artifacts produced by the Synchronization Layer are incorporated into a shared contextual representation known as the \textbf{Collective Context Field (CCF)}.

\medskip

\noindent The Collective Context Field functions as a distributed informational environment that encodes reasoning priors derived from the collective experiences of participating nodes. Each personal AI instance periodically receives updates from the CCF, which influence its reasoning processes through contextual conditioning mechanisms such as prompt priors, policy adjustments, or adapter weighting.

\medskip

\noindent In this architecture, improvements discovered by any node can propagate throughout the network via updates to the Collective Context Field. For example, if a node discovers an improved reasoning strategy for solving a particular type of problem, the synthetic learning artifacts representing this strategy can be incorporated into the shared context field. Other nodes can then incorporate this information into their reasoning processes without requiring direct model parameter updates.

\medskip

\noindent This mechanism enables knowledge propagation across the network while maintaining the autonomy of individual nodes. Each personal instance retains control over its local reasoning model and contextual state, yet benefits from the collective experiences of the network.

\medskip

\noindent Conceptually, the Collective Context Layer resembles distributed cognitive substrates in biological systems, where global modulatory signals influence neural activity across distributed networks without directly encoding detailed representations (\cite{AstonJonesCohen2005}). By separating contextual evolution from parameter updates, the architecture enables scalable distributed intelligence while avoiding the coordination challenges associated with centralized training cycles.

\subsection{Emergent Properties of the Architecture}

The interaction between the four architectural layers produces several emergent properties that distinguish the H3LIX architecture from conventional AI systems.

\medskip

\noindent First, the architecture supports \textbf{decentralized intelligence evolution}, allowing knowledge improvements to propagate across the network without requiring centralized training infrastructure. Second, it provides strong \textbf{privacy guarantees}, since raw interaction data remains local to each node and only synthetic learning artifacts are shared across the network. Third, the architecture supports \textbf{scalability through network participation}, meaning that the learning capacity of the system increases as more nodes contribute reasoning experiences. Finally, the distributed nature of the architecture provides \textbf{resilience to node failure}, since the system does not depend on a single centralized model host.

\medskip

\noindent Together, these properties enable a new class of artificial intelligence infrastructure in which distributed reasoning agents collectively contribute to the evolution of shared intelligence while preserving the autonomy and privacy of individual participants.

\section{Formalization of the Collective Context Field}
\noindent The Collective Context Field (CCF) represents the central coordination mechanism of the
H3LIX architecture. While the preceding sections described the conceptual and architectural
foundations of the system, this section formalizes the mechanism through which distributed
learning signals are aggregated and propagated across the network.

\medskip

\noindent In conventional distributed machine learning systems, improvements discovered by individual nodes are integrated into a shared global model through parameter aggregation or gradient synchronization (\cite{McMahanEtAl2017}; \cite{KairouzEtAl2021}). In contrast, the H3LIX
architecture introduces a different learning paradigm in which improvements propagate
through a shared contextual substrate rather than through direct modification of model
parameters.

\medskip

\noindent The Collective Context Field functions as a distributed informational layer that encodes
reasoning priors derived from the aggregated experiences of participating nodes. Instead of
synchronizing neural network weights, nodes exchange \textbf{synthetic learning signals} that
influence reasoning behavior through contextual conditioning.

\subsection{Network representation}
Consider a distributed network consisting of $N$ participating nodes. Each node operates a
personal AI instance that hosts a local reasoning model. Let:
\begin{itemize}
\item $\mathcal{D}_i^t \in \mathfrak{D}$ denote the private time-dependent local state at node $i$ (including, e.g., interaction history, memory, identity state, local evidence, and preferences),
\item $P_i^t \in \mathcal{P}$ denote the time-dependent local pattern object at node $i$,
\item $T\in\mathcal{Z}$ denote a task in a canonical task space $(\mathcal{Z},d_{\mathcal{Z}})$,
\item $R\in\mathcal{R}$ denote an outcome in an outcome space $\mathcal{R}$,
\item $(\mathcal{S},d_{\mathcal{S}})$ denote the shared artifact space,
\item $\Phi:\mathcal{P}\to\mathcal{A}$ and $\Psi:\mathcal{Z}\times\mathcal{R}\to\mathcal{B}$ denote canonical representation maps into metric spaces $(\mathcal{A},d_{\mathcal{A}})$ and $(\mathcal{B},d_{\mathcal{B}})$, respectively,
\item $\mathsf{Proj}:\mathcal{A}\times\mathcal{B}\to\mathcal{S}$ denote a privacy-preserving projection operator, and
\item $S_i^t \in \mathcal{S}$ denote the shareable artifact exported by node $i$ at round $t$.
\end{itemize}

\medskip

\noindent The generation of synthetic learning signals can be expressed as local task solving followed by privacy-preserving artifact construction. Given a task $T\in\mathcal{Z}$, node $i$ produces an outcome
\begin{equation*}
    R \;=\; \mathsf{Solve}\big(T;\ \mathcal{D}_i^t,\ P_i^t\big)\ \in\ \mathcal{R},
\label{eq:solve_TR_prose}
\end{equation*}
and then exports a synthetic learning signal in the form of an artifact
\begin{equation*}
S_i^{t}
\;=\;
\mathsf{Proj}\Big(\Phi(P_i^t),\ \Psi(T,R)\Big)
\ \in\ 
(\mathcal{S},d_{\mathcal{S}}).
\label{eq:artifact_construct_prose}
\end{equation*}
By construction, $S_i^t$ is a canonicalized, filtered/compressed, privacy-preserving projection of the local computation; in particular, it does not reveal the private state $\mathcal{D}_i^t$. This formulation reflects the fact that synthetic learning signals are derived from internal reasoning dynamics and observed
interaction outcomes, while raw user data remain local.

Examples of such signals include:
\begin{itemize}
\item corrected reasoning paths
\item counterfactual reasoning trajectories
\item distilled instruction--response pairs
\item reasoning policy adjustments
\end{itemize}
These signals represent the basic informational units exchanged within the distributed learning network.

\subsection{Aggregation into the Collective Context Field}
Once generated, synthetic learning signals are transmitted through the synchronization layer
and incorporated into a shared contextual representation of the network. The Collective
Context Field at time $t$ represents contextual knowledge derived from the network's
collective artifact stream.

\medskip

\noindent We define the CCF at round $t$ as the joint configuration (tuple) of the participating nodes' exported artifacts.
Let $\mathcal{I}^t\subseteq\{1,\dots,N\}$ denote the set of participating nodes at round $t$, and set
\begin{equation}
\mathcal{C}^t \;:=\; \mathcal{S}^{|\mathcal{I}^t|},\qquad
C^{\mathrm{col},t} \;:=\; (S_i^t)_{i\in\mathcal{I}^t}\ \in\ \mathcal{C}^t.
\end{equation}
This definition is well-defined under the following postulates on shareable artifacts and collective processing.

\begin{description}
\item[(U1) Canonicality.]
All exported artifacts admit a canonical representation in a common space $(\mathcal{S},d_{\mathcal{S}})$. In particular, for each
node $i$ and round $t$ the shared object $S_i^t\in\mathcal{S}$ is well-defined and comparable across nodes, enabling
collective use.

\item[(U2) Non-identifiability.]
Artifacts do not determine the underlying private state. Concretely, the mapping
\begin{equation*}
    (\mathcal{D}_i^{t},\,P_i^{t},\,T,\,R)\ \mapsto\ S_i^{t}
\end{equation*}
is many-to-one in the sense that many distinct private states and local pattern objects induce the same (or statistically
indistinguishable) artifact. This may be strengthened by requiring that $\mathsf{Proj}$ enforces differential privacy.

\item[(ND) Non-degeneracy in shared space.]
To rule out trivial encodings (e.g.\ constant artifacts), we assume that, in typical rounds $t$, the set of exported artifacts
$\{S_i^t\}_{i\in\mathcal{I}^t}$ has non-trivial spread in $(\mathcal{S},d_{\mathcal{S}})$. One convenient measure of spread is the
average squared pairwise distance
\begin{equation}
\mathrm{Disp}_t \;:=\;
\frac{1}{|\mathcal{I}^t|(|\mathcal{I}^t|-1)}\sum_{\substack{i,j\in\mathcal{I}^t\\ i\neq j}}
d_{\mathcal{S}}(S_i^t,S_j^t)^2,
\end{equation}
and we assume $\mathrm{Disp}_t$ is bounded away from $0$ on typical rounds.
\end{description}

\medskip

\noindent These postulates mean that artifacts can be meaningfully compared (U1), they do not reveal private state (U2), and the shared
channel is non-trivial (ND). Downstream collective processing of the artifact stream (e.g.\ when computing derived update signals
from $C^{\mathrm{col},t}$) may perform several functions simultaneously:
\begin{enumerate}
\item \textbf{Signal filtering}, removing inconsistent or low-confidence learning artifacts
\item \textbf{Signal weighting}, adjusting the influence of signals according to node reliability or reputation
\item \textbf{Temporal smoothing}, ensuring stability of derived collective signals over time
\item \textbf{Privacy preservation}, preventing reconstruction of individual node contributions
\end{enumerate}

%%%%%%%%%%%%

\subsection{Contextual Conditioning of Local Reasoning}
After synchronization, each node incorporates collective information derived from the Collective Context Field into its local
reasoning dynamics. In our formulation this contextual conditioning is implemented through node-side projections of the CCF and
subsequent updates of the local pattern object.

\medskip

\noindent Concretely, each node forms from the CCF a node-specific collective view
\begin{equation*}
X_i^t \;=\; \Pi_i\!\big(C^{\mathrm{col},t}\big),
\end{equation*}
where $\Pi_i$ may implement selection, relevance filtering, weighting, and privacy-preserving transformations (for example,
excluding the node's own contribution). The node then updates its local pattern object by combining private local state with this
collective experience:
\begin{equation*}
P_i^{t+1}
\;=\;
\mathsf{UpdatePat}\big(P_i^t;\ \mathcal{D}_i^{t+1},\ X_i^t\big)
\ \in\ \mathcal{P}.
\end{equation*}
The updated pattern object $P_i^{t+1}$ affects subsequent task solving through
$\mathsf{Solve}(T;\mathcal{D}_i^{t+1},P_i^{t+1})$. In application, the conditioning mechanism may influence reasoning through several pathways, including:
\begin{itemize}
\item prompt priors derived from collective signals,
\item adjustment of adapter weights or reasoning policies,
\item retrieval augmentation based on shared artifacts,
\item modification of reasoning heuristics.
\end{itemize}
Through this mechanism, improvements discovered by any node in the network can influence the reasoning behavior of other participating nodes without requiring direct modification of model parameters.

\medskip

\noindent This design allows knowledge to propagate across the network through contextual modulation rather than parameter convergence,
fundamentally distinguishing the architecture from classical federated learning systems.

%%%%%%%%%%%

\subsection{Dynamics of the Collective Context Field}
The Collective Context Field evolves continuously through the accumulation of distributed
learning signals. Several properties characterize the dynamics of this system.

\medskip

\noindent \textbf{Emergent Knowledge Propagation}

\medskip
\noindent As nodes generate synthetic learning artifacts and contribute them to the field, useful reasoning strategies gradually propagate throughout the network. Because the field aggregates signals from many nodes, improvements discovered by individual participants can influence the reasoning behavior of the entire system.

\medskip

\noindent \textbf{Noise Suppression}

\medskip

\noindent Aggregation mechanisms filter inconsistent or low-confidence signals, preventing unstable updates to the field. Weighting mechanisms based on signal agreement and node reliability further enhance the stability of the system.

\medskip

\noindent \textbf{Adaptive Learning Rate}

\medskip

\noindent The influence of newly generated learning signals can vary depending on contextual factors such as signal agreement, temporal stability, and node reputation. This dynamic weighting allows the system to adjust its effective learning rate based on the reliability of incoming information.

\medskip

\noindent \textbf{Privacy Preservation}

\medskip

\noindent Because the field operates on synthetic learning signals rather than raw interaction data, it preserves user privacy while enabling distributed collective learning.

\subsection{Comparison with Parameter-Based Distributed Learning}
The Collective Context Field introduces a fundamentally different approach to distributed
machine learning compared with traditional federated learning systems. In parameter-based
approaches, distributed nodes exchange gradients or parameter updates that are aggregated to
update a shared global model (\cite{McMahanEtAl2017}; \cite{KairouzEtAl2021}). In contrast, the
H3LIX architecture separates contextual evolution from parameter updates.

\medskip

\noindent This distinction has several important implications. First, it allows improvements in reasoning behavior to propagate across the network without requiring coordinated model retraining cycles. Second, it enables heterogeneous models to participate in the network, since contextual signals can influence reasoning behavior even if nodes operate different model architectures. Third, it may reduce communication overhead depending on artifact representation, compression, and synchronization frequency; in some cases artifacts can be smaller than full parameter updates (cf. distillation-style communication tradeoffs such as FedMD)."

\medskip

\noindent By shifting the focus of distributed learning from parameter synchronization to contextual
signal propagation, the Collective Context Field enables a form of \textbf{distributed cognitive evolution} in which knowledge emerges from the interaction of many locally adaptive
reasoning agents.

\section{Privacy-Preserving Distributed Learning Protocol}
The Collective Context Field described in the previous section provides the theoretical
mechanism through which learning signals propagate across the H3LIX network. However,
implementing such a distributed learning system requires a protocol that ensures privacy,
scalability, and robustness. This section introduces the \textbf{privacy-preserving distributed learning protocol} that enables participating nodes to contribute learning signals to the network without exposing sensitive user data or internal system states.

\medskip

\noindent In conventional centralized machine learning pipelines, training data are collected and aggregated within centralized infrastructures where model updates are computed through
large-scale gradient optimization. Federated learning systems attempt to address privacy
concerns by keeping raw data local while transmitting gradient updates to a central server
(\cite{McMahanEtAl2017}). Although this approach reduces direct exposure of training data,
gradient updates can still reveal information about underlying datasets, and large-scale
coordination of model updates introduces communication and synchronization challenges
(\cite{KairouzEtAl2021}).

\medskip

\noindent The H3LIX architecture adopts a different strategy. Instead of transmitting raw data or gradient updates, nodes exchange \textbf{synthetic learning artifacts derived from reasoning processes}. These artifacts represent distilled abstractions of reasoning improvements rather than direct reflections of the underlying interaction data. As a result, the distributed learning protocol focuses on exchanging structured knowledge signals rather than optimizing shared model parameters.

\subsection{Local Learning and Signal Generation}
The distributed learning process begins at the level of individual nodes. Each personal AI
instance continuously engages in reasoning processes while interacting with users or external
systems. During these interactions, the reasoning model generates internal reasoning traces
that capture intermediate steps used to arrive at a given output.

\medskip

\noindent After each reasoning episode, the node evaluates the outcome and extracts learning-relevant
patterns from the reasoning process. These patterns may include corrected reasoning paths,
successful reasoning strategies, policy adjustments, or generalized instruction--response
mappings. The extracted information is transformed into \textbf{synthetic learning signals}, which encode the improvement in reasoning capability without preserving the original interaction content.

\medskip

\noindent Formally, the local learning stage can be described as a transformation from interaction data
to synthetic learning artifacts:
\begin{equation*}
R \;=\; \mathsf{Solve}\big(T;\ \mathcal{D}_i^t,\ P_i^t\big)\in\mathcal{R},\qquad
S_i^{t}
\;=\;
\mathsf{Proj}\Big(\Phi(P_i^t),\ \Psi(T,R)\Big)\in\mathcal{S}.
\label{eq:local_signal_generation_aligned}
\end{equation*}
Here $\mathcal{D}_i^t$ remains private; only the artifact $S_i^t$ is transmitted.

\noindent This process ensures that sensitive user data never leaves the local node while still allowing
knowledge improvements to be shared across the network.

%%%%%%%

\subsection{Signal Validation and Filtering}

Before synthetic learning signals are transmitted to the synchronization layer, each node
performs internal validation procedures designed to ensure the quality and reliability of the
generated artifacts. These validation mechanisms evaluate signals according to criteria such as
reasoning consistency, empirical validation against known information sources, and alignment
with existing contextual priors.

\medskip

\noindent Signals that fail validation are discarded locally, preventing low-quality learning artifacts
from entering the distributed learning pipeline. In addition, nodes may assign confidence
scores to validated signals based on the reliability of the reasoning process and the consistency
of the outcome.

\medskip

\noindent These validation procedures play a critical role in maintaining the stability of the Collective
Context Field. Without such filtering mechanisms, distributed learning systems could be
vulnerable to noise accumulation or adversarial manipulation.

\medskip

\noindent In addition to node-local validation, the network may apply collective-level filtering and
weighting when \emph{using} the Collective Context Field to produce update-relevant signals.
Concretely, such processing can be modeled as part of collective improvement computation
(e.g.\ $U^t=\mathsf{Improve}(C^{\mathrm{col},t})$) and/or as part of node-specific projections
(e.g.\ $X_i^t=\Pi_i(C^{\mathrm{col},t})$), which may downweight inconsistent artifacts, enforce
robustness criteria, and incorporate privacy-preserving safeguards before influencing local
updates.

%%%%%%%

\subsection{Secure Signal Transmission}
After validation, synthetic learning artifacts are transmitted to aggregation nodes responsible
for integrating signals across the network. To ensure privacy preservation, the H3LIX
architecture employs \textbf{secure aggregation protocols} that prevent reconstruction of individual contributions from aggregated signals.

\medskip

\noindent Secure aggregation techniques allow multiple nodes to contribute encrypted learning artifacts
that can only be decoded after aggregation. As a result, individual signals remain confidential
even to the aggregation infrastructure itself. Such protocols have been widely studied in the
federated learning literature and have been shown to significantly reduce privacy risks
associated with distributed learning systems (\cite{BonawitzEtAl2017}).

\medskip

\noindent In addition to encryption-based aggregation, the system may incorporate differential privacy
mechanisms that introduce controlled noise into aggregated signals. These mechanisms
ensure that individual node contributions cannot be inferred even in the presence of auxiliary
information.

\subsection{Context Field Formation and Collective Improvement}
Once signals have been securely aggregated, the resulting information is used to form the
Collective Context Field and to compute collective improvement signals that can be propagated
back to participating nodes.

\medskip

\noindent We form the Collective Context Field at round $t$ as the tuple of participating node artifacts:
\begin{equation*}
C^{\mathrm{col},t} \;=\; (S_i^t)_{i\in\mathcal{I}^t}\ \in\ \mathcal{C}^t=\mathcal{S}^{|\mathcal{I}^t|}.
\end{equation*}
A collective improvement signal may then be computed as a derived object
\begin{equation*}
U^t \;=\; \mathsf{Improve}\big(C^{\mathrm{col},t}\big)\ \in\ \mathcal{U},
\end{equation*}
where $\mathcal{U}$ is an abstract improvement space (e.g.\ shared priors, curricula, or update directives).

\medskip

\noindent The signal $U^t$ may be disseminated to nodes, and nodes may also form node-specific projections
\begin{equation*}
X_i^t=\Pi_i(C^{\mathrm{col},t})
\end{equation*}
to incorporate collective information into local adaptation. In particular, local pattern objects may be
updated via
\begin{equation*}
P_i^{t+1}=\mathsf{UpdatePat}(P_i^t;\mathcal{D}_i^{t+1},X_i^t),
\end{equation*}
allowing improvements discovered anywhere in the network to influence local behavior without direct parameter synchronization.

\subsection{Periodic Model Consolidation}
Although the H3LIX architecture primarily relies on contextual signal propagation rather than
direct parameter updates, periodic consolidation events may occur in which accumulated
contextual improvements are distilled into updated model parameters. These consolidation
phases allow the system to translate persistent contextual patterns into more efficient model
representations.

\medskip

\noindent During consolidation, aggregated learning artifacts stored in the Collective Context Field are
used to generate training datasets that reflect the accumulated knowledge of the network.
These datasets can be used to fine-tune base reasoning models or to train new model versions
that incorporate the improvements discovered through distributed learning.

\medskip

\noindent Importantly, such consolidation events occur \textbf{periodically rather than continuously},
allowing the system to benefit from distributed contextual learning without requiring constant
retraining of centralized models.

\subsection{Security and Robustness Considerations}
Distributed learning systems must be resilient to malicious or unreliable participants. The
H3LIX architecture therefore incorporates several mechanisms designed to enhance network
robustness. These mechanisms include signal validation procedures, node reputation systems,
anomaly detection algorithms, and cross-node verification of learning artifacts.

\medskip

\noindent Reputation mechanisms assign reliability scores to participating nodes based on the historical
consistency of their contributions. Nodes that consistently generate high-quality signals
receive greater influence in the aggregation process, while unreliable nodes gradually lose
influence.

\medskip

\noindent Anomaly detection algorithms further monitor the distributed learning pipeline for unusual
patterns that may indicate adversarial activity or corrupted signals. Through these
mechanisms, the system aims to maintain stability and trustworthiness even in large
heterogeneous networks.

\subsection{Implications for Distributed AI Learning}
The distributed learning protocol described in this section enables the H3LIX architecture to
combine several desirable properties rarely achieved simultaneously in large-scale AI
systems. These include privacy preservation, decentralized knowledge generation, and
continuous collective learning.

\medskip

\noindent By relying on synthetic learning signals rather than raw data or gradient updates, the
architecture allows nodes to contribute knowledge to the network without exposing sensitive
information. At the same time, the use of contextual signal propagation enables improvements
in reasoning capability to spread across the network without requiring large centralized
training runs.

\medskip

\noindent These properties suggest a new model for distributed AI systems in which intelligence
evolves through the interaction of many locally adaptive agents operating within a shared
contextual environment.

%%%%%%%%%%%%%%%%%%%%%%%%%%%%%%%%%%%

\section{Energy-Adaptive Model Evolution}
The continued expansion of large-scale artificial intelligence systems has increasingly exposed the energy constraints associated with modern machine learning infrastructure. Training frontier-scale neural networks requires substantial computational resources and prolonged operation of high-performance accelerators. As model sizes and training datasets continue to grow, energy consumption has become a critical factor influencing both the economic and environmental sustainability of AI development (\cite{StrubellGaneshMcCallum2019}).

\medskip

\noindent Recent studies examining the environmental impact of deep learning have shown that training large neural models can produce significant carbon emissions when energy is derived from fossil-fuel-based electricity grids. These concerns have motivated increasing interest in energy-efficient machine learning methods and carbon-aware computing strategies designed to reduce the environmental footprint of AI systems (\cite{PattersonEtAl2021}). Building on carbon-aware computing and renewable-aware workload scheduling (\cite{PattersonEtAl2021}; \cite{RadovanovicEtAl2023}; \cite{ToosiEtAl2017}), Energy-Adaptive Model Evolution treats learning and synchronization as deferrable workloads whose timing (and potentially participation) is modulated by real-time energy/carbon/price signals. Under favorable conditions (temporal flexibility, accurate forecasts, and sufficient renewable availability), this can reduce marginal emissions and shift energy-intensive operations away from high-carbon periods, though total footprint still scales with total compute.

\medskip

\noindent The H3LIX architecture introduces an alternative perspective through the concept of \textbf{Energy-Adaptive Model Evolution (EAME)}. Instead of treating energy consumption as an external constraint imposed on AI systems, the architecture integrates energy availability directly into the dynamics of distributed learning. In this framework, the timing and intensity of learning processes across the network can adapt to real-time energy conditions, allowing AI computation to align with renewable energy availability and grid conditions.

\subsection{Energy Constraints in Frontier AI Systems}
Large-scale neural network training requires sustained operation of specialized hardware such as graphics processing units (GPUs) or tensor processing units (TPUs). These accelerators consume substantial electrical power when operating at full capacity, and large training runs often involve thousands of such devices operating simultaneously for extended periods.

\medskip

\noindent As a result, the energy requirements of frontier AI models have become a growing concern within both the machine learning community and the broader technology sector. Studies analyzing the carbon footprint of natural language processing models have shown that training large transformer models can require considerable computational energy, particularly when training pipelines are executed within data centers powered by carbon-intensive electricity sources (\cite{StrubellGaneshMcCallum2019}).

\medskip

\noindent While improvements in hardware efficiency and training algorithms have partially mitigated these concerns, the overall trajectory of model scaling suggests that energy demand will remain a critical limiting factor. Consequently, future AI architectures must increasingly consider energy availability as a fundamental design constraint.

\subsection{Renewable Energy Variability and Computational Scheduling}
One of the defining characteristics of renewable energy sources such as solar and wind power is their \textbf{temporal variability}. Energy production from these sources fluctuates according to weather conditions, seasonal cycles, and geographic factors. As renewable energy adoption increases globally, electricity grids must increasingly balance variable energy supply with fluctuating demand.

\medskip

\noindent Recent research has explored the potential of \textbf{carbon-aware computing}, in which computational workloads are scheduled dynamically based on the carbon intensity of electricity generation (\cite{PattersonEtAl2021}). In such systems, non-time-critical computing tasks can be shifted to periods when renewable energy availability is high or when electricity grids have surplus capacity.

\medskip

\noindent The distributed nature of the H3LIX architecture makes it particularly well suited to this type of adaptive scheduling. Because the network consists of many independent nodes performing local computation, learning tasks can be distributed across time and geography according to energy availability. Instead of relying on continuous operation of large centralized clusters, the system can perform intensive learning activities during periods of abundant renewable energy.

\subsection{Energy-Adaptive Learning Cycles}
Energy-Adaptive Model Evolution introduces a mechanism through which distributed learning activities across the H3LIX network can be scheduled according to real-time energy conditions. In this framework, nodes participating in the network monitor local energy signals provided by energy management systems (EMS), electricity markets, or renewable energy forecasting systems.

\medskip

\noindent When renewable energy availability is high or electricity demand is low, nodes may increase computational activity to perform tasks such as synthetic signal generation, model refinement, or synchronization operations. Conversely, during periods of limited energy availability, nodes may reduce computational activity and focus primarily on inference and interaction tasks.

\medskip

\noindent This adaptive scheduling mechanism allows the network to perform energy-intensive learning operations during periods when the environmental and economic cost of computation is lowest. Because learning processes are distributed across many nodes and occur asynchronously, the network can accumulate learning improvements over time without requiring continuous high-intensity computation.

\subsection{Integration with Distributed Learning Architecture}
The integration of energy-adaptive scheduling with the distributed learning architecture described in previous sections creates a system in which learning dynamics can evolve in response to both informational and environmental signals. The Synthetic Learning Layer generates learning artifacts based on reasoning processes, while the Synchronization Layer coordinates the exchange of these artifacts across the network. Energy-adaptive scheduling influences when and how frequently these processes occur.

\medskip

\noindent For example, nodes may generate synthetic learning signals continuously during interaction with users, but large-scale aggregation and contextual field updates may occur primarily during periods when sufficient energy resources are available. Similarly, periodic model consolidation events described in Section 6 can be scheduled during periods of renewable energy surplus, minimizing the environmental impact of training operations.

\medskip

\noindent This integration allows the architecture to decouple the \textbf{temporal distribution of learning activity from user interaction patterns}, enabling computationally intensive operations to occur at times when energy costs are lowest.

\subsection{Implications for Sustainable AI Infrastructure}
Energy-Adaptive Model Evolution introduces a new perspective on the relationship between artificial intelligence systems and energy infrastructure. Rather than treating computational workloads as fixed processes that require continuous energy supply, the H3LIX architecture treats learning activities as \textbf{adaptive processes that can align with environmental conditions}.

\medskip

\noindent This approach has several potential implications. First, it may enable the development of AI infrastructures that scale without proportionally increasing their environmental footprint. Second, it creates opportunities for integrating AI computation with emerging renewable energy ecosystems, particularly in regions where surplus renewable energy generation would otherwise be curtailed. Third, the distributed nature of the architecture allows computation to occur closer to energy production sites, reducing the need for centralized data center infrastructure.

\medskip

\noindent By coupling distributed learning processes with energy availability, the architecture suggests a pathway toward AI systems that evolve in harmony with sustainable energy infrastructure rather than competing with it for scarce resources.

\section{Scaling Properties of the Architecture}
The success of modern artificial intelligence systems has been closely tied to the concept of \textbf{scaling laws}, which describe how model performance improves as a function of increased computational resources, model parameters, and training data. Empirical research has demonstrated that large neural language models exhibit predictable improvements in performance as model size and training compute increase (\cite{KaplanEtAl2020}). Subsequent work has refined these observations, showing that optimal performance depends on balancing model size, dataset size, and computational budget in what has been described as compute-optimal scaling (\cite{HoffmannEtAl2022}).

\medskip

\noindent While these scaling laws have enabled remarkable progress in artificial intelligence capabilities, they have also created a strong dependency on centralized computational infrastructure. Training frontier models now requires extensive clusters of specialized hardware and significant energy resources, concentrating AI development within a small number of organizations capable of operating hyperscale data centers. This concentration raises concerns regarding both the sustainability and accessibility of future AI development.

\medskip

\noindent The H3LIX architecture proposes an alternative scaling pathway based on \textbf{distributed contextual learning across a network of autonomous AI instances}. Instead of scaling intelligence primarily through increases in model parameters or centralized training compute, the architecture scales through the participation of distributed nodes contributing synthetic learning signals to a shared contextual environment. This section examines the scaling properties of this architecture and compares them with existing scaling paradigms in machine learning.

\subsection{Parameter Scaling in Centralized AI Systems}
The dominant scaling paradigm in modern AI development involves increasing model capacity through larger neural networks trained on larger datasets. Empirical studies have shown that language model performance improves approximately as a power-law function of model parameters, dataset size, and training compute (\cite{KaplanEtAl2020}). These scaling relationships have motivated the development of increasingly large models trained using massive datasets and computational infrastructure.

\medskip

\noindent Subsequent research has refined the understanding of optimal scaling by demonstrating that many models are trained inefficiently when model size grows faster than dataset size or training compute. The compute-optimal scaling framework proposed by Hoffmann et al. (2022) suggests that optimal performance requires balancing these factors to avoid over- or under-training models relative to their parameter count.

\medskip

\noindent Although parameter scaling has produced substantial improvements in AI performance, it introduces several structural limitations. First, the computational requirements for training frontier models grow rapidly as model size increases, making large-scale training accessible only to organizations with significant computational resources. Second, training large models requires massive datasets, which are increasingly difficult to obtain as publicly available text corpora become saturated. Third, centralized training pipelines concentrate both data and computational resources within large data centers, raising concerns about privacy, governance, and environmental impact.

\medskip

\noindent These limitations have motivated increasing interest in alternative scaling approaches that can extend the capabilities of AI systems without relying exclusively on larger centralized models.

\subsection{Distributed Training and Federated Learning}
Distributed machine learning approaches represent one alternative strategy for scaling AI systems. In distributed training architectures, model training is distributed across multiple computing nodes that collaborate to optimize a shared model. Such approaches are commonly used in large-scale deep learning systems where model training is parallelized across clusters of GPUs or specialized accelerators (\cite{DeanEtAl2012}).

\medskip

\noindent Federated learning represents a more decentralized variant of distributed training in which individual devices collaboratively train a shared model without sharing raw training data (\cite{McMahanEtAl2017}). In federated learning systems, each participating device computes local model updates using its own data, and these updates are aggregated to produce a global model. This approach has been widely studied as a privacy-preserving alternative to centralized data collection and training (\cite{KairouzEtAl2021}).

\medskip

\noindent Despite their advantages, both distributed training and federated learning remain fundamentally grounded in the \textbf{parameter-centric paradigm of machine learning}. In these systems, distributed nodes collaborate to update a shared set of model parameters through gradient-based optimization. As a result, the ultimate objective of distributed training remains convergence toward a unified global model.

\medskip

\noindent Recent work has explored extensions to federated learning that support heterogeneous models or parameter-efficient training methods such as LoRA-based updates (\cite{HuEtAl2021}). Other research has investigated peer-to-peer learning architectures in which nodes exchange model updates without centralized coordination. However, these approaches still assume that collective intelligence ultimately emerges through parameter synchronization across nodes.

\subsection{Contextual Scaling Through Collective Learning}
The H3LIX architecture introduces a fundamentally different scaling mechanism in which intelligence evolves through the propagation of \textbf{contextual learning signals} rather than through parameter synchronization. In this paradigm, each node contributes synthetic learning artifacts derived from local reasoning processes. These artifacts are aggregated into the Collective Context Field, which conditions reasoning behavior across the network.

\medskip

\noindent The scaling properties of the system can therefore be expressed in terms of network participation rather than model parameters. Let $N$ denote the number of participating nodes and $S_i$ denote the learning signals generated by node $i$. The total learning activity of the network can be approximated as:
$L(t)=\sum_{i\in\mathcal{I}^t}\ell(S_i^t)$,
where $L$ represents the total volume of learning signals generated across the network.

\medskip

\noindent As the number of participating nodes increases, the system accumulates a larger and more diverse set of learning experiences. Because these experiences are distilled into contextual learning artifacts rather than raw data or gradient updates, they can be integrated into the Collective Context Field without requiring synchronization of model parameters.

\medskip

\noindent In this architecture, improvements discovered by any node can influence the reasoning behavior of other nodes through updates to the contextual field. Consequently, the effective learning capacity of the network grows with the number of participating nodes and the diversity of reasoning experiences they generate.

\medskip

\noindent This scaling mechanism resembles forms of collective intelligence observed in distributed human knowledge systems such as open-source software communities and collaborative knowledge platforms. In such systems, collective knowledge evolves through the aggregation of contributions from many independent participants rather than through centralized optimization processes.

\subsection{Comparison of Scaling Paradigms}
The differences between the three primary scaling paradigms discussed above can be summarized conceptually.

\medskip

\noindent In \textbf{centralized scaling}, improvements in model capability are achieved primarily by increasing model parameters, training data, and computational resources. In \textbf{distributed parameter training}, improvements emerge from collaborative optimization of a shared model across multiple nodes. In contrast, \textbf{contextual scaling}, as proposed in the H3LIX architecture, enables improvements to propagate through the exchange of learning artifacts and contextual signals.
These paradigms differ not only in their technical mechanisms but also in their implications for system governance, infrastructure requirements, and scalability. Centralized scaling concentrates computational power within large data centers, while contextual scaling distributes learning processes across many independent nodes.
This shift from parameter-centric learning to contextual signal propagation introduces a new scaling variable for AI systems: \textbf{network participation}. As more nodes contribute learning artifacts to the network, the diversity of reasoning experiences increases, enabling the Collective Context Field to encode increasingly rich contextual knowledge.

\subsection{Implications for Future AI Development}
The contextual scaling paradigm introduced by the H3LIX architecture suggests that future AI systems may evolve through mechanisms resembling distributed cognitive ecosystems rather than monolithic neural networks. In such systems, intelligence emerges from the interaction of many locally adaptive agents operating within a shared informational environment.

\medskip

\noindent This perspective aligns with broader research trends emphasizing modular, agent-based, and distributed AI architectures. Recent work in multi-agent reinforcement learning and collaborative learning systems has demonstrated that networks of interacting agents can exhibit emergent problem-solving capabilities that exceed the capabilities of individual agents operating in isolation (\cite{BusoniuBabuskaDeSchutter2008}).

\medskip

\noindent If contextual learning mechanisms such as the Collective Context Field can be implemented effectively, they may provide a pathway toward scalable AI systems that combine the advantages of distributed intelligence with the powerful reasoning capabilities of modern neural models. Such systems could potentially overcome some of the structural limitations associated with centralized model scaling while enabling more collaborative and decentralized forms of artificial intelligence development.

\section{Roadmap to Frontier Capability}
The preceding sections have introduced the conceptual and architectural foundations of the H3LIX system and demonstrated how distributed contextual learning may offer an alternative scaling paradigm for artificial intelligence. However, a critical question remains: how could a distributed architecture such as H3LIX realistically achieve capabilities comparable to contemporary frontier models?

\medskip

\noindent Large-scale language models currently achieve their performance primarily through centralized training pipelines involving massive datasets and extensive computational resources (\cite{KaplanEtAl2020}; \cite{HoffmannEtAl2022}). Any alternative architecture seeking to compete with these systems must provide a credible pathway for reaching similar levels of capability while leveraging its distributed learning mechanisms.

\medskip

\noindent This section outlines a phased roadmap through which the H3LIX architecture could progressively evolve from an initial distributed AI system into a large-scale collective intelligence infrastructure. Each phase builds upon the capabilities of the previous phase while introducing new mechanisms that increase the learning capacity and coordination efficiency of the network.

\subsection{Phase I: Personal AI Instance Deployment}
The first phase of the roadmap focuses on establishing the foundational infrastructure of the H3LIX network through the deployment of \textbf{personal AI instances}. In this phase, each participating node operates a locally hosted AI system capable of performing conversational reasoning, contextual analysis, and task assistance.

\medskip

\noindent Rather than attempting to train a frontier-scale model from scratch, early deployments would rely on existing language models that are optimized for local execution through techniques such as parameter-efficient fine-tuning and model compression (\cite{HuEtAl2021}). These models would serve as reasoning engines embedded within the broader cognitive architecture described in previous sections.

\medskip

\noindent During this phase, the primary objective is to establish persistent identity continuity and contextual memory within individual nodes. Personal AI instances would accumulate interaction histories, develop user-specific contextual knowledge, and begin generating synthetic learning signals derived from reasoning processes. Although learning would remain largely local during this stage, nodes would begin producing structured learning artifacts that could later contribute to distributed learning.

\medskip

\noindent This phase lays the foundation for a decentralized AI ecosystem in which users interact with persistent AI instances that operate independently of centralized model hosting infrastructures.

\subsection{Phase II: Distributed Adapter Specialization}
In the second phase, the architecture introduces mechanisms for \textbf{distributed model specialization} through adapter-based learning modules. Parameter-efficient techniques such as low-rank adaptation (LoRA) allow neural models to incorporate task-specific capabilities without modifying the underlying base model parameters (\cite{HuEtAl2021}).

\medskip

\noindent Within the H3LIX architecture, these adapter modules can be trained locally on individual nodes based on interaction patterns and domain-specific tasks. Over time, different nodes within the network may develop specialized adapters tailored to particular domains, such as scientific reasoning, legal analysis, or technical problem solving.

\medskip

\noindent The distributed nature of the network allows these adapters to evolve independently across nodes while still contributing to the collective knowledge base through synthetic learning signals. As a result, the system gradually develops a distributed ecosystem of specialized reasoning capabilities.

\medskip

\noindent During this phase, the network begins to accumulate diverse reasoning experiences across nodes, increasing the richness of the learning signals available for aggregation within the Collective Context Field.

\subsection{Phase III: Collective Context Amplification}
The third phase introduces large-scale synchronization of learning signals across the network through the \textbf{Collective Context Field}. At this stage, nodes begin regularly exchanging synthetic learning artifacts derived from local reasoning processes.

\medskip

\noindent These artifacts are aggregated and incorporated into the contextual field, allowing reasoning improvements discovered by individual nodes to propagate throughout the network. Over time, the field accumulates a large body of contextual knowledge derived from diverse reasoning experiences.

\medskip

\noindent The Collective Context Field functions as a shared cognitive substrate that influences reasoning behavior across the network. Nodes receiving contextual updates incorporate these signals into their reasoning processes through prompt conditioning, retrieval mechanisms, or adapter adjustments.

\medskip

\noindent This phase marks the transition from isolated personal AI systems to a \textbf{distributed collective intelligence network}, in which learning experiences generated by one node can influence the reasoning capabilities of others.

\subsection{Phase IV: Synthetic Self-Training Loops}
Once the network has accumulated a sufficiently large body of contextual knowledge, the architecture can begin leveraging \textbf{synthetic self-training loops} to accelerate learning. In this phase, nodes actively generate synthetic reasoning tasks designed to test and refine reasoning capabilities.

\medskip

\noindent For example, nodes may simulate problem-solving scenarios, generate counterfactual reasoning challenges, or engage in collaborative reasoning exchanges with other nodes. These activities produce additional learning artifacts that expand the contextual knowledge available within the Collective Context Field.

\medskip

\noindent Self-generated training data has already been shown to play an important role in modern AI development. Techniques such as self-play and synthetic dataset generation have been used to train systems capable of achieving superhuman performance in complex tasks (\cite{SilverEtAl2017}). By incorporating similar mechanisms within a distributed network of reasoning agents, the H3LIX architecture can potentially generate large volumes of high-quality training signals without relying exclusively on external datasets.

\medskip

\noindent During this phase, the network begins to exhibit properties of \textbf{autonomous collective learning}, where the system continuously generates new learning opportunities through simulated reasoning activities.

\subsection{Phase V: Periodic Frontier Model Consolidation}
The final phase of the roadmap involves periodic consolidation of accumulated contextual knowledge into updated base models. While the architecture relies primarily on contextual signal propagation for continuous learning, periodic consolidation events allow the system to incorporate long-term learning patterns into improved model parameters.

\medskip

\noindent During consolidation, synthetic learning artifacts stored in the Collective Context Field are transformed into training datasets that reflect the collective reasoning experiences of the network. These datasets can be used to fine-tune base reasoning models or train new model versions that incorporate the knowledge accumulated through distributed learning.

\medskip

\noindent Importantly, these consolidation events occur periodically rather than continuously, reducing the need for constant centralized training. Instead, large-scale model updates occur only after the network has accumulated sufficient new knowledge to justify retraining.

\medskip

\noindent This hybrid approach combines the advantages of distributed contextual learning with the performance improvements achievable through parameter optimization. Over time, the system can produce increasingly capable reasoning models while maintaining the decentralized learning infrastructure that enables continuous knowledge evolution.

\subsection{Long-Term Evolution of Distributed Intelligence}
The phased roadmap outlined above illustrates how a distributed architecture such as H3LIX could progressively evolve toward frontier-level AI capability. Rather than attempting to compete directly with centralized training pipelines from the outset, the architecture gradually accumulates knowledge through distributed reasoning experiences and synthetic learning signals.

\medskip

\noindent As the network grows and more nodes participate in collective learning, the diversity and volume of reasoning experiences increase. This expansion enhances the richness of the Collective Context Field and enables more effective knowledge propagation across the network.

\medskip

\noindent In the long term, such a system could support a form of distributed artificial intelligence infrastructure in which the capabilities of the network emerge from the coordinated activity of many locally adaptive agents. By combining distributed learning mechanisms with periodic model consolidation, the architecture offers a pathway toward scalable AI systems that evolve through continuous collective learning rather than through isolated training cycles.

\section{Theoretical Implications}
The H3LIX architecture introduces a distributed framework for artificial intelligence in which intelligence evolves through the interaction of many locally operating AI instances coordinated by a shared contextual substrate. Beyond its technical architecture, the system has broader theoretical implications for the study of artificial intelligence systems, distributed cognition, and the governance of complex digital infrastructures. This section discusses three principal theoretical contributions: the reconceptualization of large-scale AI systems as distributed cognitive infrastructures, the introduction of contextual learning as an alternative scaling paradigm, and the implications of human–AI symbiosis for the design of future intelligent systems.

\subsection{Artificial Intelligence as Distributed Cognitive Infrastructure}
Contemporary artificial intelligence systems are often conceptualized as centralized computational artifacts, typically represented by a single large neural model trained on large datasets. In this paradigm, intelligence is localized within the model itself, and improvements in capability are achieved primarily through scaling model parameters and training data (\cite{KaplanEtAl2020}; \cite{HoffmannEtAl2022}).

\medskip

\noindent The architecture proposed in this paper challenges this assumption by introducing a system in which intelligence emerges from the \textbf{interaction of multiple AI agents embedded within a distributed learning network}. In such a system, reasoning capability is not confined to a single model but arises from the coordinated activity of many locally adaptive nodes connected through the Collective Context Field.

\medskip

\noindent This perspective aligns with broader theories of \textbf{distributed cognition}, which propose that cognitive processes are not confined to individual agents but emerge from interactions between agents, tools, and informational environments (\cite{Hutchins1995}). Distributed cognition research has shown that complex reasoning tasks in human systems often arise from collaborative processes involving multiple individuals and artifacts operating within shared information environments.

\medskip

\noindent The H3LIX architecture extends this idea to artificial intelligence systems by creating a distributed environment in which AI instances collectively generate and share reasoning signals. Rather than viewing intelligence as a property of a single model, the architecture treats intelligence as an \textbf{emergent property of the network itself}.

\medskip

\noindent This reconceptualization of AI systems as distributed cognitive infrastructures may have significant implications for the design of future AI ecosystems. It suggests that the capabilities of AI systems may increasingly depend not only on the capabilities of individual models but also on the \textbf{structure and dynamics of the networks in which those models operate}.

\subsection{Contextual Learning as an Alternative Scaling Paradigm}
The second theoretical contribution concerns the introduction of \textbf{contextual learning as a scaling mechanism} for artificial intelligence. Current research in machine learning has largely focused on scaling model parameters, training data, and computational resources as the primary drivers of performance improvements (\cite{KaplanEtAl2020}; \cite{HoffmannEtAl2022}).

\medskip

\noindent While this paradigm has been highly successful, it has also produced systems that require increasingly large computational resources and centralized infrastructure. The H3LIX architecture proposes an alternative mechanism in which learning improvements propagate through contextual signals rather than through parameter updates.

\medskip

\noindent In this paradigm, nodes within the network generate synthetic learning signals derived from reasoning processes. These signals are aggregated into the Collective Context Field and subsequently influence reasoning behavior across the network. The result is a system in which improvements discovered by individual nodes can propagate through contextual conditioning rather than through direct parameter synchronization.

\medskip

\noindent This approach introduces a new scaling variable for AI systems: \textbf{network participation}. As more nodes participate in the network and contribute learning signals, the diversity and volume of reasoning experiences increase. The Collective Context Field integrates these experiences into a shared contextual representation that can influence reasoning behavior across the network.

\medskip

\noindent This scaling mechanism resembles knowledge accumulation processes observed in collaborative human knowledge systems such as open-source software communities or scientific research networks. In these systems, collective knowledge evolves through contributions from many independent participants rather than through centralized optimization processes.

\medskip

\noindent If contextual learning mechanisms can be implemented effectively, they may enable AI systems to scale through \textbf{collective experience accumulation} rather than through continued expansion of model parameters and centralized training datasets.

\subsection{Human–AI Symbiosis and Cognitive Co-Evolution}
The third theoretical implication concerns the relationship between humans and artificial intelligence systems. Many current AI systems are designed as tools that provide assistance in specific tasks but do not maintain persistent contextual relationships with users. In contrast, the H3LIX architecture introduces the concept of \textbf{persistent personal AI instances} that interact continuously with individual users over extended periods of time.

\medskip

\noindent Such systems have the potential to form long-term collaborative relationships with users in which both the human and the AI instance influence each other’s learning processes. The AI instance accumulates contextual knowledge about the user’s preferences, goals, and reasoning patterns, while the user adapts their behavior based on the capabilities and suggestions provided by the AI system.

\medskip

\noindent This form of interaction resembles what has been described in cognitive science as \textbf{symbiotic cognition}, in which human reasoning processes become intertwined with external cognitive artifacts such as tools, written language, or digital systems (\cite{ClarkChalmers1998}). In such systems, cognitive processes are distributed across biological and technological components that interact continuously.

\medskip

\noindent By embedding personal AI instances within distributed learning networks, the H3LIX architecture creates the possibility for a new form of human–AI symbiosis in which individuals participate in a collective intelligence system while maintaining control over their personal AI environments. Users benefit from the knowledge accumulated across the network while contributing their own reasoning experiences to the collective learning process.

\medskip

\noindent This model of AI development suggests that future intelligent systems may evolve not as isolated computational tools but as \textbf{co-evolving socio-technical systems} in which humans and AI agents participate jointly in the creation and refinement of knowledge.

\subsection{Implications for AI Governance and Decentralized Intelligence}
The distributed architecture proposed in this paper also raises important questions regarding the governance of artificial intelligence systems. Centralized AI infrastructures concentrate significant decision-making power within a small number of organizations that control training data, computational resources, and model deployment.

\medskip

\noindent By contrast, distributed architectures such as H3LIX distribute intelligence generation across many nodes operated by independent participants. In such systems, governance mechanisms must address issues such as signal validation, trust, and reputation within the network.

\medskip

\noindent Reputation-based aggregation mechanisms, anomaly detection systems, and distributed consensus protocols may play an important role in ensuring the stability and reliability of distributed learning networks. These mechanisms can help prevent the propagation of low-quality or adversarial signals while preserving the decentralized nature of the system.

\medskip

\noindent From a broader perspective, distributed AI architectures may enable more participatory models of AI development in which users contribute directly to the evolution of intelligent systems. Such systems could support forms of \textbf{collective intelligence governance} in which decision-making processes are distributed across the network rather than concentrated within centralized institutions.

\section{Practical Applications and Use Cases}
The distributed architecture proposed in this paper enables a range of potential applications that extend beyond the capabilities of conventional centralized AI systems. Because the H3LIX framework combines persistent personal AI instances with distributed contextual learning, it supports environments in which intelligence emerges from the interaction of many autonomous nodes operating within a shared informational substrate.

\medskip

\noindent This section outlines several domains in which such a distributed cognitive infrastructure may offer significant advantages, including personal AI systems, distributed research networks, collaborative knowledge ecosystems, and large-scale decision-support infrastructures.

\subsection{Persistent Personal AI Systems}
One of the most immediate applications of the H3LIX architecture is the development of \textbf{persistent personal AI systems} capable of maintaining long-term contextual relationships with individual users. Unlike conventional conversational AI systems that operate through stateless interactions with centralized models, personal AI instances within the H3LIX network maintain persistent contextual knowledge about their users.

\medskip

\noindent Such systems can support a wide range of activities including knowledge management, decision assistance, personal productivity support, and long-term learning. Because contextual information is stored locally within each personal instance, users retain control over their personal data while still benefiting from knowledge accumulated across the broader network.

\medskip

\noindent Over time, personal AI instances could evolve into sophisticated cognitive assistants that support complex reasoning tasks such as research synthesis, planning, and strategic analysis. Through the Collective Context Field, these instances can incorporate insights derived from distributed learning experiences without requiring centralized storage of user data.

\subsection{Distributed Scientific and Research Collaboration}
Another promising application domain lies in \textbf{distributed research environments}, where scientists, engineers, and researchers collaborate across institutional and geographic boundaries. Research processes often involve the integration of diverse knowledge sources and reasoning perspectives, making them well suited for distributed cognitive systems.

\medskip

\noindent Within such environments, personal AI instances could assist researchers by organizing literature, synthesizing experimental results, and identifying emerging patterns in scientific data. Learning signals generated during these activities could contribute to the Collective Context Field, allowing insights discovered by one research group to influence reasoning processes across the broader network.

\medskip

\noindent This distributed approach could accelerate knowledge discovery by enabling collaborative learning across institutions while preserving control over sensitive research data. Instead of sharing raw datasets, researchers could contribute distilled reasoning artifacts that represent validated insights or methodological improvements.

\medskip

\noindent Such architectures may support new forms of \textbf{collective scientific intelligence}, in which the reasoning capabilities of individual researchers are augmented by distributed AI systems capable of integrating insights from across the global research community.

\subsection{Collaborative Knowledge Systems}
The H3LIX architecture also provides a foundation for \textbf{collaborative knowledge ecosystems} that extend beyond traditional information-sharing platforms. Systems such as open-source software communities and collaborative knowledge repositories already demonstrate the power of distributed knowledge generation.

\medskip

\noindent By embedding AI agents within these environments, the architecture enables the creation of dynamic knowledge systems in which both humans and AI instances contribute to the evolution of shared understanding. AI agents operating within these systems could assist with tasks such as verifying information sources, synthesizing conflicting perspectives, and identifying gaps in collective knowledge.

\medskip

\noindent Through the Collective Context Field, improvements in reasoning strategies or knowledge representations discovered within one part of the network could propagate across the entire system. This capability may enable collaborative knowledge systems that evolve more rapidly and robustly than traditional centralized knowledge repositories.

\subsection{Organizational and Decision-Support Systems}
Organizations increasingly rely on complex decision-making processes that require integrating large volumes of information across multiple domains. Distributed AI architectures such as H3LIX could support \textbf{organizational decision-support systems} capable of synthesizing insights from multiple teams, departments, or external data sources.

\medskip

\noindent Personal AI instances associated with different organizational units could generate reasoning signals derived from local decision-making processes. These signals could contribute to the Collective Context Field, allowing the organization to accumulate institutional knowledge in a distributed and privacy-preserving manner.

\medskip

\noindent Such systems could support strategic planning, risk analysis, and operational coordination across complex organizations. Because learning signals are derived from reasoning processes rather than raw data, sensitive organizational information can remain localized while still contributing to collective intelligence formation.

\subsection{Resilient Distributed AI Infrastructure}
The decentralized nature of the H3LIX architecture also offers potential advantages for the development of \textbf{resilient AI infrastructures} capable of operating across geographically distributed environments. In contrast to centralized AI services that depend on large data centers, distributed networks of AI instances can continue operating even if individual nodes become unavailable.

\medskip

\noindent This property may be particularly valuable in contexts where reliable centralized infrastructure is difficult to maintain. For example, distributed AI networks could support communication, information analysis, and decision-support functions in environments characterized by intermittent connectivity or limited centralized computing resources.

\medskip

\noindent In strategic or security-sensitive contexts, distributed intelligence systems may also enhance the robustness of information processing infrastructures. Because intelligence generation is distributed across many nodes, the system can continue functioning even if portions of the network are disrupted. Such architectures could therefore support resilient decision-support capabilities in complex operational environments, including humanitarian coordination, emergency response, or defense-related analytical systems.

\medskip

\noindent While the architecture described in this paper is primarily intended for civilian and collaborative applications, the underlying principles of distributed contextual learning may also have relevance for \textbf{strategic analysis, situational awareness systems, and multi-agent coordination environments}, where distributed reasoning capabilities are advantageous.

\subsection{Human–AI Symbiotic Ecosystems}
Finally, the most transformative application of the H3LIX architecture may lie in the development of \textbf{human–AI symbiotic ecosystems}. In such environments, humans interact continuously with personal AI instances that participate in distributed learning networks. Over time, these interactions create feedback loops in which both human reasoning and AI capabilities evolve together.

\medskip

\noindent Human users contribute contextual insights and domain expertise, while AI instances assist with information synthesis, reasoning augmentation, and knowledge integration. The Collective Context Field allows these interactions to contribute to the evolution of a shared cognitive environment in which knowledge and reasoning strategies accumulate over time.

\medskip

\noindent Such systems may enable new forms of collective intelligence in which human and artificial agents collaborate to address complex challenges that exceed the capabilities of individual actors. By combining distributed AI learning with persistent human–AI interaction, the architecture provides a potential foundation for future socio-technical systems in which intelligence emerges from the interaction of humans and machines operating within shared informational ecosystems.

\section{Conclusion}
This paper presented the \textbf{H3LIX Decentralized Frontier Model Architecture (DFMA)}, a systems-level proposal for distributed AI in which multiple locally operating instances contribute to collective improvement through a shared informational layer. The motivation is the set of constraints increasingly associated with centralized large-model development---compute concentration, energy cost, data governance, and limited personalization---highlighted in prior work on scaling and training regimes \parencite{KaplanEtAl2020,HoffmannEtAl2022} and environmental impacts \parencite{StrubellGaneshMcCallum2019,PattersonEtAl2021}.

\medskip

\noindent At the core of the proposed learning mechanism is a unified formalization of \emph{what can be shared} and \emph{how it is used}. Each node solves tasks locally via a private state and a local pattern object, producing outcomes that are distilled into shareable artifacts through canonical representation maps and a privacy-preserving projection operator. The shared channel is modeled as a common artifact space $(\mathcal{S},d_{\mathcal{S}})$, together with explicit postulates: (U1) \emph{canonicality} (comparability across nodes), (U2) \emph{non-identifiability} (non-invertibility of the mapping from private state to artifacts, optionally strengthened by secure aggregation and differential privacy), and (ND) \emph{non-degeneracy} (the shared channel must remain informative and influence collective dynamics). This provides a general interface for artifact-based collective learning that is compatible with established privacy-preserving distributed learning techniques such as secure aggregation \parencite{BonawitzEtAl2017}.

\medskip

\noindent The DFMA learning loop is naturally expressed as a two-level process: (i) node-level adaptation driven by private interaction state and local patterns, and (ii) collective processing that aggregates artifacts to compute improvement signals which condition subsequent local behavior. In this sense, H3LIX is best viewed as a unifying architectural framework that can incorporate and extend existing distributed learning paradigms by treating ``shared context'' as a first-class, structured object defined in a canonical space.

\medskip

\noindent The paper also discussed two system integrations. First, separating reasoning engines from identity, evidence tracking, and execution governance provides a modular design stance consistent with broader trends in agentic AI architectures \parencite{LewisEtAl2020,SchickEtAl2023,YaoEtAl2023,PackerEtAl2023MemGPT}, while emphasizing continuity and safety constraints at the system level. Second, \textbf{Energy-Adaptive Model Evolution} positions learning and synchronization workloads as potentially deferrable operations that may be scheduled in a carbon-/energy-aware manner \parencite{PattersonEtAl2021,RadovanovicEtAl2023,ToosiEtAl2017}, expressing these ideas as part of the distributed learning loop rather than as an external deployment optimization.

\medskip

\noindent Finally, several aspects of the architecture admit useful analogies to distributed cognition and biological learning (e.g., global modulation and replay-like simulation). These analogies are offered as interpretive motivation rather than as claims of biological equivalence \parencite{AstonJonesCohen2005,WilsonMcNaughton1994,FosterWilson2006,Hutchins1995}. The next step is empirical validation: evaluating artifact leakage risks under explicit threat models, quantifying the effect of artifact conditioning on downstream performance, and assessing stability/robustness under heterogeneous and adversarial nodes \parencite{KairouzEtAl2021}.

\medskip

\noindent More broadly, the H3LIX perspective suggests that future AI systems may evolve not only through increasingly large centralized models, but also through networks of locally adaptive agents operating within shared informational environments. The DFMA framework represents one possible pathway toward such systems and clarifies the assumptions and interfaces required to make artifact-based collective learning feasible at scale.

\printbibliography

\end{document}